# Routing *Physarum* with electrical flow/current

Soichiro Tsuda[1], Jeff Jones[2], Andrew Adamatzky[2], Jonathan Mills[3]

[1] Exploratory Research for Advanced Technology (ERATO),
Japan Science and Technology Agency,
Yamadaoka 1-5, Suita, Osaka 565-0871, Japan
tsuda-soichiro@bio.eng.osaka-u.ac.jp

[2] Centre for Unconventional Computing,
University of the West of England, Bristol, BS16 1QY, UK
jeff.jones@uwe.ac.uk, andrew.adamatzky@uwe.ac.uk

[3] School of Informatics and Computing, Indiana University Bloomington,
901 E. 10th St., Bloomington, IN 47408, USA
jwmills@cs.indiana.edu

**Abstract:** Plasmodium stage of *Physarum polycephalum* behaves as a distributed dynamical pattern formation mechanism who's foraging and migration is influenced by local stimuli from a wide range of attractants and repellents. Complex protoplasmic tube network structures are formed as a result, which serve as efficient `circuits' by which nutrients are distributed to all parts of the organism. We investigate whether this `bottom-up' circuit routing method may be harnessed in a controllable manner as a possible alternative to conventional template-based circuit design. We interfaced the plasmodium of *Physarum polycephalum* to the planar surface of the spatially represented computing device, (Mills' Extended Analog Computer, or EAC), implemented as a sheet of analog computing material whose behaviour is input and read by a regular 5x5 array of electrodes. We presented a pattern of current distribution to the array and found that we were able to select the directional migration of the plasmodium growth front by exploiting plasmodium electro-taxis towards current sinks. We utilised this directional guidance phenomenon to route the plasmodium across its habitat and were able to guide the migration around obstacles represented by repellent current sources. We replicated these findings in a collective particle model of *Physarum polycephalum* which suggests further methods to orient, route, confine and release the plasmodium using spatial patterns of current sources and sinks. These findings demonstrate proof of concept in the low-level dynamical routing for biologically implemented circuit design.

INTRODUCTION

The plasmodium stage of the giant single-celled true slime mould *Physarum polycephalum* is demonstrated by a complex morphological interaction with its environment. The growth and migration of the amorphous acellular organism is affected by the presence of both attractants and repellents. The plasmodium forms a spatially efficient protoplasmic tube transport network, within which nutrients are transported within the mass of the organism. The efficiency of this network is a trade-off in terms of material cost, path redundancy and resilience to damage. The topography of the transport network may be viewed as a living `circuit' pattern by which is constructed in a bottom-up manner in a system comprising relatively simple components, guided by local stimuli. This contrasts with classical top-down methods of circuit design which utilise globally applied templates to enforce connectivity patterns between components.

Growing *Physarum* circuits can be controlled by various means (A. Adamatzky, 2010b), including physical modification of growth substrate (Figure 1a), chemo-repellents (A. Adamatzky) (Figure 1b) or even high-concentration of chemo-attractants (Figure 1c). When oat flakes colonised by plasmodium are placed in the northern part of a 12x12cm Petri dish filled with agar, *Physarum* cells are attracted by virgin oat flakes in the southern part of the dish. Physically removed substrate, sodium chloride, and sugar in high concentration all act as repellents. Also, as non-contact methods of geometrically shaped illumination (A. Adamatzky, 2009) (Figure 2). In this case, white light acts as repellents and 8-shaped window is cut in the aluminium foil underlying the Petri dish. The 8-shaped window is illuminated by Seiko El Sheet (See details of the experiments in (A. Adamatzky, 2009)). Using of the techniques modifying physical structure of substrate or its chemical components would make slime mould circuits non-reusable, because chemical species continue to diffuse in the substrate. Moreover, even the case of simple removal of substrate would not be suitable because they do not allow for reconfiguration of the circuits. Non-invasive control techniques, as e.g. with light (A. Adamatzky, 2009) would be appropriate, however they are rather unreliable and do not always predictable control of the growing slime mould.

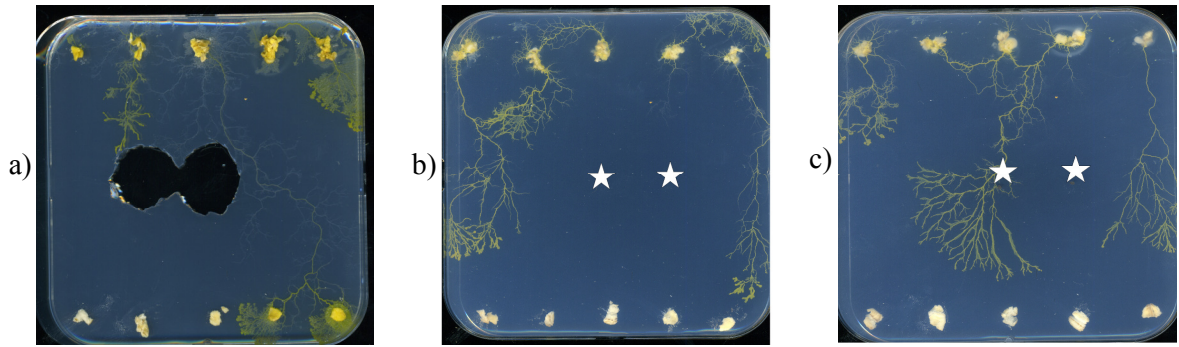

**Figure 1:** Controlling *Physarum* by "conventional" means.
(a) Physical removal of substrate: 8-shaped piece of agar plate removed is visible as black shape.
(b) Sodium chloride: position of two grains of salt is shown by stars.
(c) High concentration of sugar: several grains of sugar are placed in the positions indicated by stars. Images of Petri dishes are scanned on Epson Perfection 4490. Images are taken 24h after inoculation.}

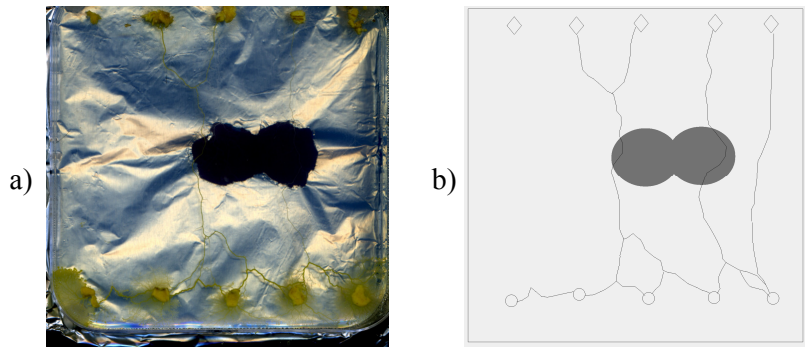

**Figure 2:** Controlling *Physarum* by "conventional" means. Oat flakes colonised by plasmodium and virgin (oat flakes are positioned similarly as in Figure 1)
(a) Photograph of the experimental Petri dish (under-dish illumination is switched off) is done with FujiFilm Fine-Pix digital camera.
(b) Scheme of the experiment: seeds of *Physarum* are shown by rhombs, destination oaf flakes by discs, major protoplasmic tubes reflecting trajectories of propagating acting zones by lines. }

In this paper we investigate the possible exploitation of biologically derived circuit pattern construction by interfacing the plasmodium stage of *Physarum polycephalum* with the behaviour of a spatially implemented general purpose analog computer, the EAC.

EXPERIMENTS

The *Physarum* plasmodium is known to shown negative electrotaxis, a behaviour that the cell tends to migrate towards the direction of cathode under an electric field (Anderson, 1951). Although the physiological mechanism of this behaviour is barely known (Anderson, 1962), other types of motile cells are also known to show negative electrotaxis (Adler & Shi, 1988; Cooper & Schliwa, 1986). Considering that the *Physarum* plasmodium exhibits other various tactic behaviour (e.g. temperature, chemical, white light), this is one of information processing capabilities in slime moulds in order to survive under complex environment. In this section, we employ this tactic behaviour of the cell to investigate the possibility of *Physarum* circuit construction using the Extended Analog Computer (EAC) developed by Jonathan Mills and co-workers (Mills et al., 2006).

EXPERIMENT SETUP

The plasmodium of *Physarum polycephalum* was cultured on the 1.5% de-ionised agar gel (Sigma-Aldrich, UK) without any nutrients in a culture dish (245x245x25mm$^3$, Nunclon™ Surface, Nunc, USA). Culture dishes were kept in the dark at 25˚C and fed with oat flakes once or twice a day. A medium on which *Physarum* cells migrate and an electric field is applied is agar gel. In particular, 1.5% tap-water agar gel was used because the medium needs to be conductive and also because it was used in the original literature (Anderson, 1951). A rectangle-shaped agar (approximately 10x10x1cm$^3$) was cut out of a larger agar gel and placed on 5x5 electrode matrix of the EAC. The agar and EAC were separated by a plastic transparency sheet with 25 tiny punched holes so that electrode pins will stick out but any undesired short-circuits because of the conductive agar can be avoided. The EAC, originally developed as a modern re-interpretation of traditional analog computer, provides flexible control of spatial current distribution using 5x5 electrode matrix. We used this device to form various spatial current distributions in the conductive agar. It was controlled from a PC via RS232 (serial) connection using a custom software program written in Java.

BEHAVIOUR OF SLIME MOULD UNDER UNIFORM ELECTRIC FIELD

We first investigated if *Physarum* plasmodia show the negative phototactic behaviour on the EAC. Eight pieces of *Physarum* cells are taken from a large cell culture (approximately 100mg each) and placed on the locations indicated as ``X'' in Figure 3a. The experiment was repeated twice (in total 16 samples). Each electrode in the rightmost column is used as a source of 50μA DC current and the leftmost electrodes are used as sinks (-50μA each). This creates a spatial electric current distribution pattern in the conductive agar gel, as shown in Figure 3b. We found the middle electrode in the source electrodes has a hardware failure and therefore the current around the electrode was low. However, a reasonably uniform gradient of electric current distribution was observed in other areas of the gel. The EAC board with *Physarum* cells were kept in the dark at room temperature throughout the experiment.

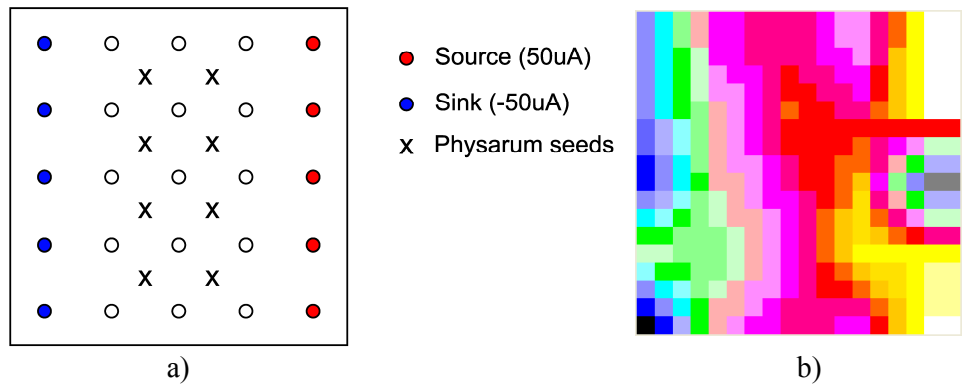

**Figure 3:** (a) A schematic diagram of EAC electrode configuration. Currents are applied from five electrodes in the rightmost column (red circles) and another five electrodes in the leftmost column are used as current sinks (blue circles). (b) A spatial distribution of electric current formed in the conductive agar gel (as measured by the custom EAC control software).

Figure 4 shows a result of the electrotaxis experiment. Eight *Physarum* seeds taken from a large cell culture (approximately 100mg each) are placed in the centre of the medium (Figure 4a) left untouched for several hours. Figure 4b shows a result of an experiment after six hours. Cells developed several branches to the surroundings. Electrotaxis of a *Physarum* cell was determined by comparing areas of the cell growing towards cathode (left-hand side) and anode (right-hand side) with the site of cell inoculation as the origin. If the former area is larger than the latter, the cell was considered to have shown the negative electrotaxis. In total, 13 out of 16 samples (81.25%) showed negative phototaxis. The remaining 3 samples did not show positive electrotaxis (moving to the right-hand side). They all tended to develop branches downwards (e.g. the bottom two cells of the right column in Figure 4b). The migration direction never changed throughout the experiment. For example, if a cell starts to grow towards cathode, it did not switch towards anode but always remained to grow in the same direction. This fact supports the electrotactic behaviour on the EAC because, under no stimulus condition, *Physarum* cells on an agar gel develop branches towards almost all the directions and the growth polarity frequently changes. More importantly, 81.25\% is a reasonably high probability in biological experiments. Thus, we concluded that the *Physarum* plasmodium shows negative electrotaxis on the conductive tap-water agar using the EAC.

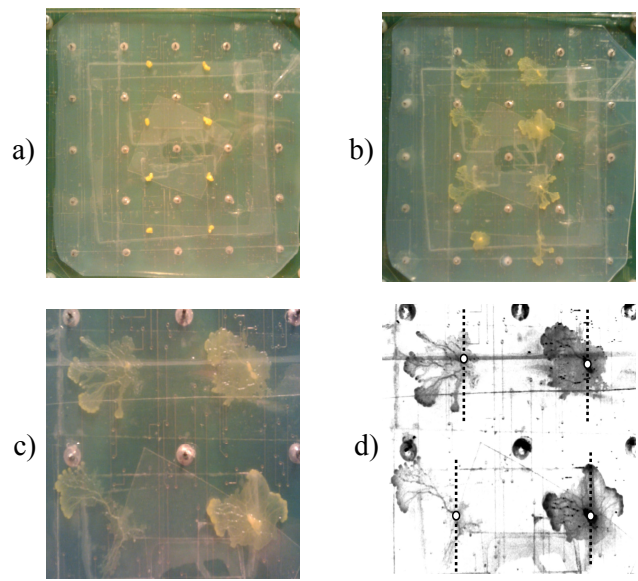

**Figure 4:** (a) A photo of the EAC board just after *Physarum* cells are inoculated on the conductive agar gel. (b) The growth of cells six hours after the inoculation. (c) A closer image of (b). Top four cells are shown. (d) A processed image of (c). White circles indicate the sites where *Physarum* cells are inoculated. The sizes of a cell growing to the left and right of the central line (dotted) are compared.

GUIDING SLIME MOULD BEHAVIOUR WITH THE SPATIAL DISTRIBUTION OF ELECTRIC FIELD

Having confirmed that *Physarum* cells show the negative electrotactic behaviour on the EAC, next we investigated the guiding control of the cell migration by a non-uniform electric field. In the previous section, the spatial distribution of the electric field was uniform, i.e. five electrodes on each side were used as current source and sinks, respectively. The electric current gradient was thus formed one-dimensionally. For the guiding control experiment, additional two electrodes with 20µA current are used as current sources (Figure 5a). Two sink electrodes were modified accordingly in order to satisfy Kirchhoff's law (to balance out the incoming and outgoing currents to the agar gel). Accordingly the spatial distribution of electric currents on a 1.5\% tap-water agar gel using this configuration was modified as in Figure 5b. It has a high current spot in the top centre area because of the newly added current sources, although it also has a small low current spot due to the hardware failure, as mentioned in the previous section. Six pieces of *Physarum* cells (approximately 100mg each) are placed at the locations indicated as "X" in Figure 5a. The EAC board with cells was kept in the condition same as the electrotaxis experiment in the previous section (dark and room temperature).

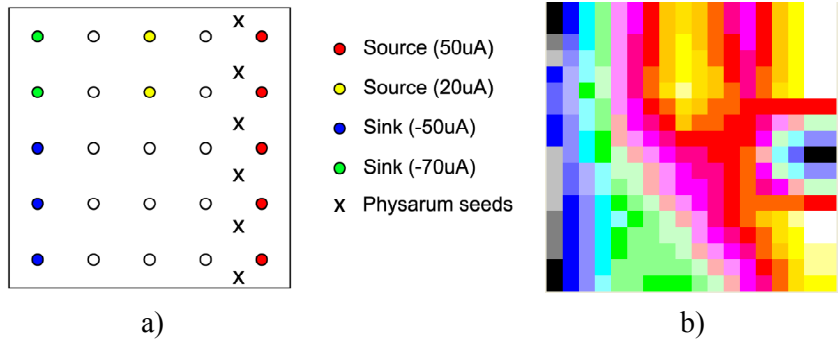

a)                                                 b)

**Figure 5:** (a) A schematic diagram of EAC electrode configuration. Two additional electrodes (yellow circles) are used as current sources. Top two sink electrodes (green circles) are changed to -70µA accordingly. (b) A spatial distribution of electric current in this configuration.

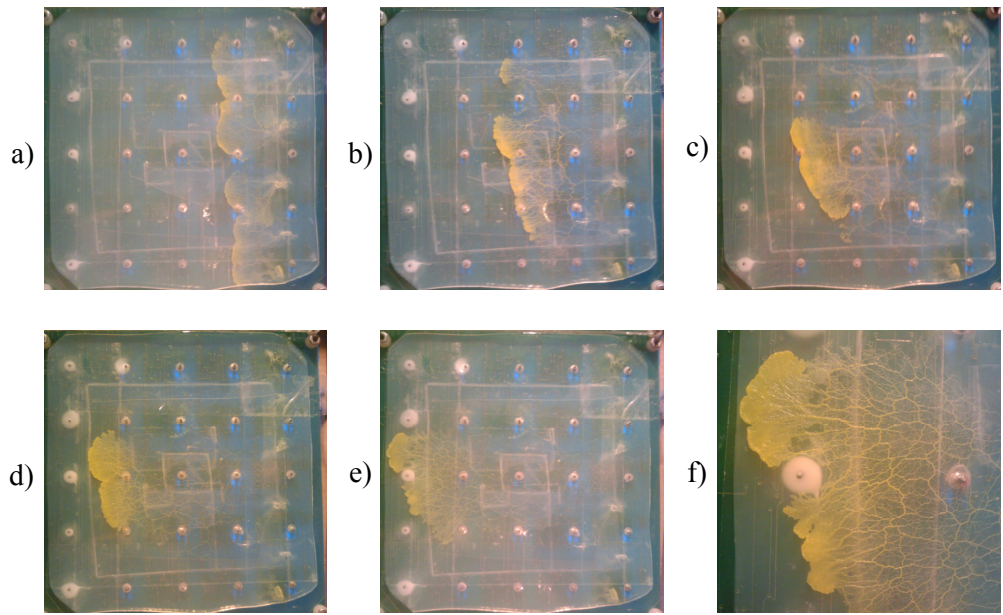

**Figure 6:** Snapshots of a guiding control experiment of slime mould.
(a) After 8 hours from the start of the experiment, (b) After 11 hours, (c) After 16 hours, (d) After 18 hours, (e) After 21 hours, (f) A closer image of (e). See text for details.

Figure 6 shows snapshots of a growth of *Physarum* cells. Within a couple of hours after the inoculation of cells, they started to grow on the agar gel and migrate towards the sink electrodes. They merged together with neighbouring cells and eventually fused into one single cell after eight hours (Figure 6a). When the merged cell reached the middle column of the electrode array, it tended to avoid the additional source electrodes and growing front split into two (Figure 6b). After 16 hours from the start, the smaller front growing between additional source electrodes disappeared (Figure 6c) due to the negative phototaxis. After that, the growing front kept growing forward (Figure 6d) and eventually arrived at the sink electrode region (Figure 6e). It should be noted that the *Physarum* cell did not touch the sink electrodes, but just grew around it (Figure 6f). This is possibly because electrolytes were formed on the sink electrodes (e.g. white circles in Figure 6, e and f), which cells do not like (In general, *Physarum* plasmodia do not like salts (Durham & Ridgway, 1976)).

MODELLING

OREGONATOR-BASED MODEL OF ELECTROTACTIC BEHAVIOUR

Profile of plasmodium's active zone growing on a non-nutrient substrate is isomorphic to shapes of wave-fragments in sub-excitable media (A Adamatzky, De Lacy Costello, & Shirakawa, 2008). When active zone of *P. polycephalum* spreads, two processes occur simultaneously --- advancing of the wave-shaped tip of the pseudopodium and formation of the trail of protoplasmic tubes. We simulate the chemo-tactic travelling of plasmodium using two-variable Oregonator equations (Field & Noyes, 1974; Tyson & Fife, 1980):

$$\frac{\partial u}{\partial t} = \frac{1}{\varepsilon}(u - u^2 - (fv + \phi)\frac{u-q}{u+q}) + D_u \nabla^2 u$$

$$\frac{\partial v}{\partial t} = u - v$$

The variable $u$ is abstracted as a local density of plasmodium's protoplasm and $v$ reflects local concentration of metabolites and nutrients. We integrate the system using Euler method with five-node Laplace operator, time step $\Delta t = 5 \cdot 10^{-3}$ and grid point spacing $\Delta x = 0.25$, with the following parameters: $\phi = \phi_0 - \eta/2$, A=0.0011109, $\phi_0$=0.0766 for imitating plasmodium growth on a non-nutrient substrate and $\phi_0$=0.066 for nutrient-substrate, $\varepsilon$=0.03, $f$=1.4, $q$=0.022.

Parameters $q$ and $f$ are inherited from the Oregonator model of Belousov-Zhabotinsky medium (Field & Noyes, 1974; Tyson & Fife, 1980), $\phi$ is proportional to local concentration of attractants and repellents. The parameter $\eta$ corresponds to a gradient of chemo-attractants emitted by data planar points. Let **P** be a set of attraction sites **P** and $x$ be a site of a simulated medium then $\eta_x = 2 \cdot 10^{-2}$ - $\min_{p \varepsilon \mathbf{P}} \{d(x, p) : \gamma(p) = \text{TRUE}\} \cdot b^{-1}$ where $3.1 \cdot 10^2 \leq b \leq 4.9 \cdot 10^2$ and $d(x, p)$ (for the simulated medium 400 x 400 sites) is an Euclidean distance between sites $x$ and $p$.

The repelling effect of source electrodes on the way of plasmodium propagation was imitated as follows. Let $x$ be a centre of a tablet radius $r > 0$, we assume $\phi_0$=0.065 in any site $y$ such that $r \geq d(y, x) \leq r + w$, where $w'$ is a the feeding zone's width. The medium is perturbed by an initial excitation, where 11x11 sites are assigned $u$=1.0 each. The perturbation generates a propagation wave-fragment travelling along gradient $\eta$. The model is proved to be successful in simulating various aspects of *Physarum* behaviour, see e.g. (A. Adamatzky, 2010b).

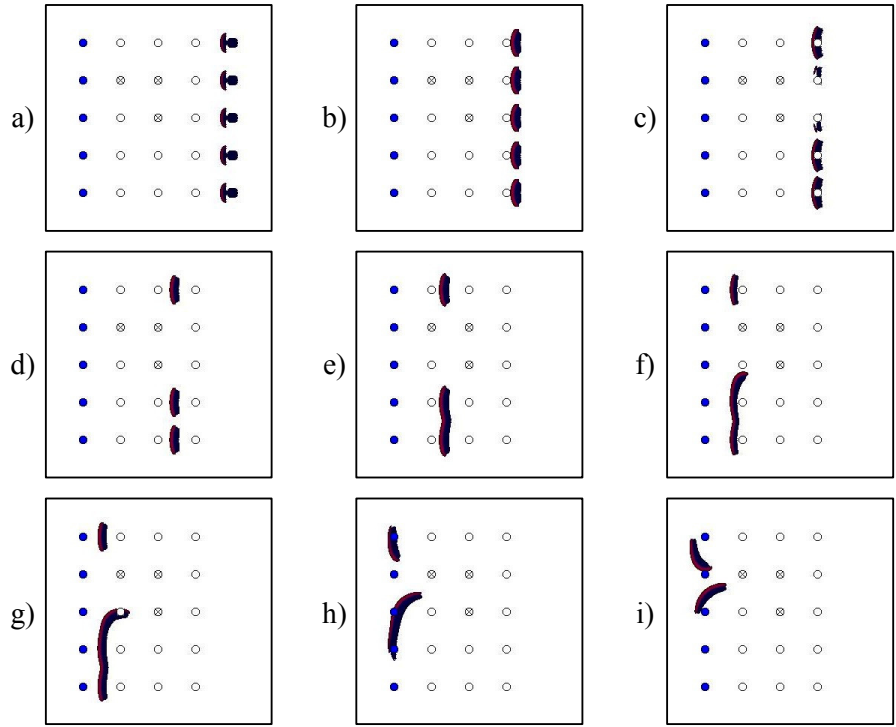

**Figure 7:** Simulating propagating plasmodia in Oregonator model. a-i) Snapshots of Oregonator model of plasmodium propagating on electrode array at $t$ = 200, 600, 800, 1200, 2000, 2400, 2600, 3000 and 3200 steps of numerical integration respectively. Sink electrodes are blue/grey discs, "obstacle"-source electrodes are shown by circles with diagonal crosses.

Excitation waves fragments are initiate at the position of plasmodium seeds, at the eastern part of array (Figure 7a) The wave-fragments travel towards the sink electrodes at the western part (Figure 7, b, c). The wave-fragments encountering "obstacle"-source electrodes either annihilate or change their shape and direction of propagation (Figure 7, d-i).

To imitate formation of the protoplasmic tubes we store values of $u$ in matrix **L**, which is processed at the end of simulation. For any site $x$ and time step $t$ if $u_x > 0.1$ and $\mathbf{L}_x=0$ then $\mathbf{L}_x=1$. The matrix **L** represents time lapse superposition of propagating wave-fronts (Figure 8a). The simulation is considered completed when propagating pattern envelops **P** and halts any further motion. At the end of simulation we repeatedly apply the erosion operation (A. Adamatzky, 2010b), which represents a stretch-activation effect (Kamiya, 1959) necessary for formation of plasmodium tubes, to **L**. The resultant protoplasmic network (Figure 8b) provides a good phenomenological match for networks recorded in laboratory experiments. The Oregonator-based model qualitatively well matches experimental results (Figure 7).

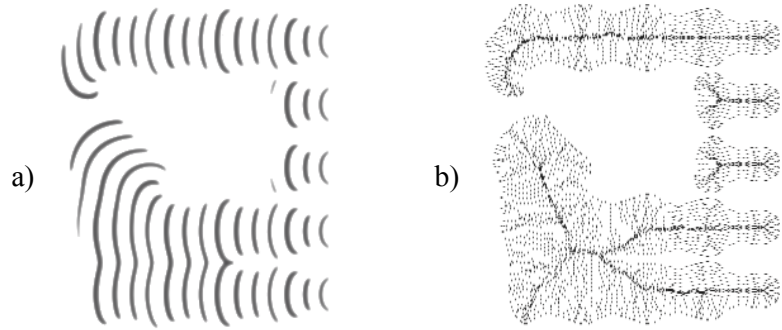

**Figure 8:** Simulating propagating plasmodia in Oregonator model.
(a) Time-lapsed images of the active zones
(b) Imitation of major protoplasmic tubes developed on an array of electrodes

PARTICLE-BASED MODEL OF ELECTROTACTIC BEHAVIOUR

To approximate the electrotactic behaviour of the plasmodium we employ and extend the particle model introduced in (J Jones, 2010) which was used to generate dynamical emergent transport networks. The approach uses a population of mobile particles with very simple behaviours, residing within a 2D diffusive lattice. The lattice (where the features of the environment are mapped to grey-scale values in a 2D image) stores particle positions and the concentration of a local factor which we refer to generically as chemoattractant. The 'chemoattractant' factor actually represents the hypothetical flux of sol within the plasmodium. Free particle movement represents the sol phase of the plasmodium. Particle positions represent the fixed gel structure (i.e. global pattern) of the plasmodium. The particles act independently and iteration of the particle population is performed randomly to avoid any artefacts from sequential ordering. The behaviour of the particles occurs in two distinct stages, the sensory stage and the motor stage. In the sensory stage, the particles sample their local environment using three forward biased sensors whose angle from the forwards position (the sensor angle parameter, SA), and distance (sensor offset, SO) may be parametrically adjusted (Figure 9a). The offset sensors represent the overlapping and intertwining filaments and generate local coupling of sensory inputs and movement to represent the protoplasmic tubes forming the transport networks and the mass of the plasmodium itself, (Figure 9, c and d respectively). The overlapping sensors correspond to the coupling caused by the cross-linking of actin filaments in motile cells (Pollard & Borisy, 2003). The SO distance is measured in pixels and a minimum distance of 3 pixels is required for strong local coupling to occur. During the sensory stage each particle changes its orientation to rotate (via the parameter rotation angle, RA) towards the strongest local source of chemoattractant (Figure 9b). After the sensory stage, each particle executes the motor stage and attempts to move forwards in its current orientation (an angle from 0-360°) by a single pixel forwards. Each lattice site may only store a single particle and – critically – particles deposit chemoattractant into the lattice only in the event of a successful forwards movement (Figure 10a). If the next chosen site is already occupied by another particle the default (i.e. non-oscillatory) behaviour is to abandon the move and select a new random direction (Figure 10b). We use the oscillatory behaviour motor condition to establish persistent forward movement, whereby if the next site is blocked a random change in direction is not selected and the particle continues to face in the current direction and increments an internal position counter until a space becomes free (Figure 10c). This mode of motor behaviour generates collective and distributed oscillations within the particle collective and was shown to replicate the emergence of oscillatory behaviour in the plasmodium and the transition between different types of oscillation patterns in an enclosed well (Tsuda & Jones, 2011).

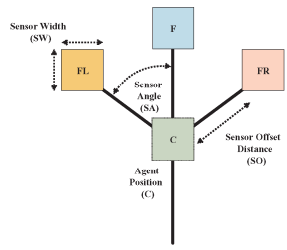

```
[Sensory stage]

- Sample chemoattractant map values
- if (F > FL) && (F > FR)
      - Continue facing same direction
- Else if (F < FL) && (F < FR)
      Rotate by RA towards larger of FL and FR
- Else if (FL < FR)
      Rotate right by RA
- Else if (FR < FL)
      Rotate left by RA
- Else
      Continue facing same direction
```

(a)                          (b)

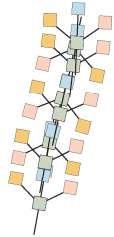 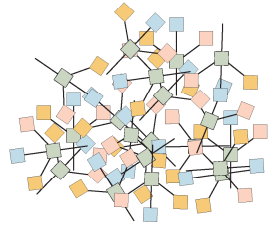

(c)                          (d)

**Figure 9:** Particle morphology and schematic illustration of overlapping particle positions representing transport networks and plasmodium mesh.
(a) Morphology showing agent position 'C' and sensor positions (FL, F, FR)
(b) Algorithm for particle sensory stage
(c) Transport network formation
(d) Overlapping sensors representing plasmodium mesh.

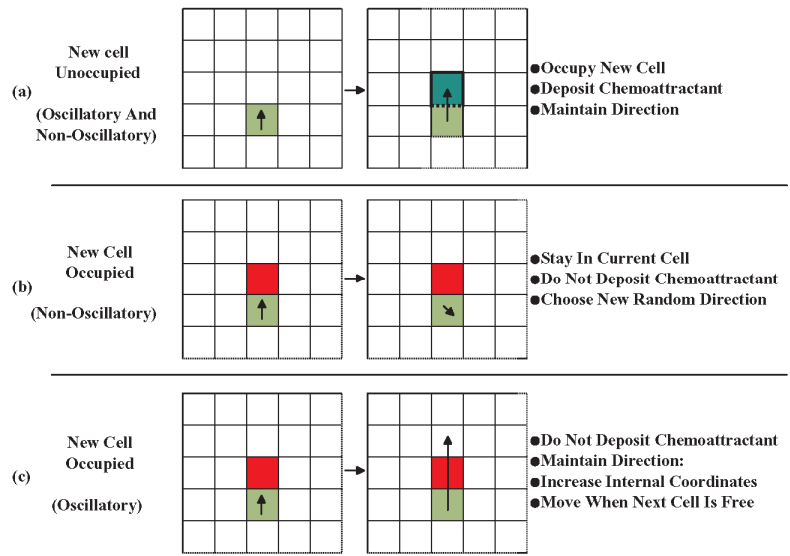

**Figure 10:** Particle motor behaviour in non-oscillatory and oscillatory modes.
(a) Behaviour in both modes is identical when new site is unoccupied
(b) When the new site is occupied in non-oscillatory mode a new random direction is selected
(c) When the new site is occupied in oscillatory mode the particle increments an internal position counter at every subsequent motor step until a new site in the current direction becomes free.

The current gradient established by the electrodes of the EAC is approximated by the 2D diffusion field of the particle model. Diffusion within the field is achieved via a simple mean filter kernel centred about each cell in the lattice. During each diffusion step every pixel in the lattice diffusion field is replaced with $(1 - d)\ u$, where $u$ is the mean of a 15x15 pixel window surrounding the current cell and $d$ is a damping parameter to limit diffusion distance. The diffusion field is initialised at zero. Current source inputs are represented by projecting negative values (-255) into the field at every scheduler step and current sinks are represented by projecting positive values (255) into the field at every step. When the field is stabilised (by diffusing the field for 400 scheduler steps) a gradient is established between sources and sinks.

An agent particle deposits 5 units of 'chemoattractant' into the diffusion field after each successful forward movement. Agent deposition into the diffusion field is significantly lower than the gradient created by the distortion of the field by the current sinks and sources. Agent sensor parameters were set to: SA=22.5, RA=45 and SO=5 and a static population size was used to reflect the experimental conditions of a non-nutrient agar substrate.

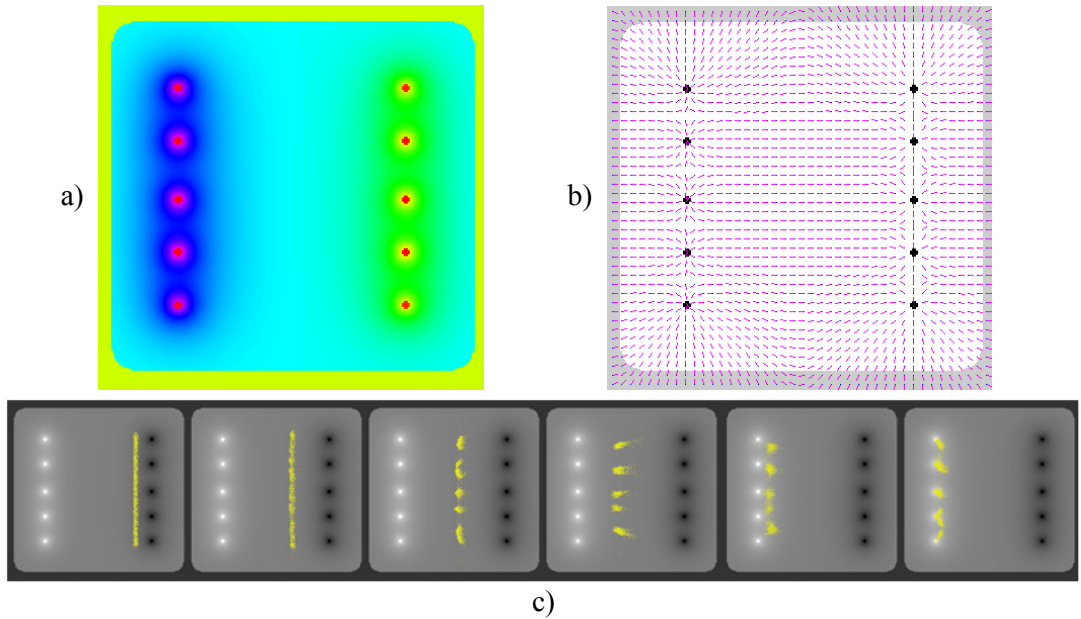

**Figure 11:** Particle approximation of plasmodium migration to current sinks.
(a) 2D representation of current gradient between current sources (right) and sinks (left)
(b) Vector field representation of current gradient field
(c) Migration of particle collective towards current sink electrodes. Image snapshots taken at 235, 285, 315, 360, 425 and 450 scheduler steps (bottom)

We began by establishing whether collective particle movement was affected by the simulated current gradient in the diffusion field. A 2D visualisation of a stable gradient field is shown in (Figure 11a) with 5 vertical current sources on the right side, and 5 vertical current sinks on the left side.
Figure 11b shows a vector field representation of the current gradient illustrating the field paths away from the current sources towards the current sinks.

The particle population (fixed size, 1500 particles) was initialised in a vertical line, just to the left of the current sources, representing the experimental stage where individual plasmodium seeds have merged. The collective then migrated as a single entity towards the current sinks (Figure 11c). The uniform migration of the population was disrupted somewhat as the migration continued. The collective, unlike the plasmodium, adhered to the sink electrodes when they were reached, as electrolytic effects at the electrodes are beyond the scope of the model approximation.

To represent obstacle avoidance during electrotaxis the particle population was again was initialised in a vertical line, just to the left of the current sources. The obstacles were represented by additional current sources in the middle of the environment resulting in a more complex gradient field and vector paths (Figure 12a,b). The collective then migrated towards the current sinks (Figure 12c) but avoided the obstacles in the middle of the field by deforming the trajectory of the collective. This deformation guided the lower part of the population around the obstacles and a smaller fragment separated and migrated to the top of the field away from the source nodes. The separation of the paths was due to the upper region being in a region of repulsion from the surrounding current sources.

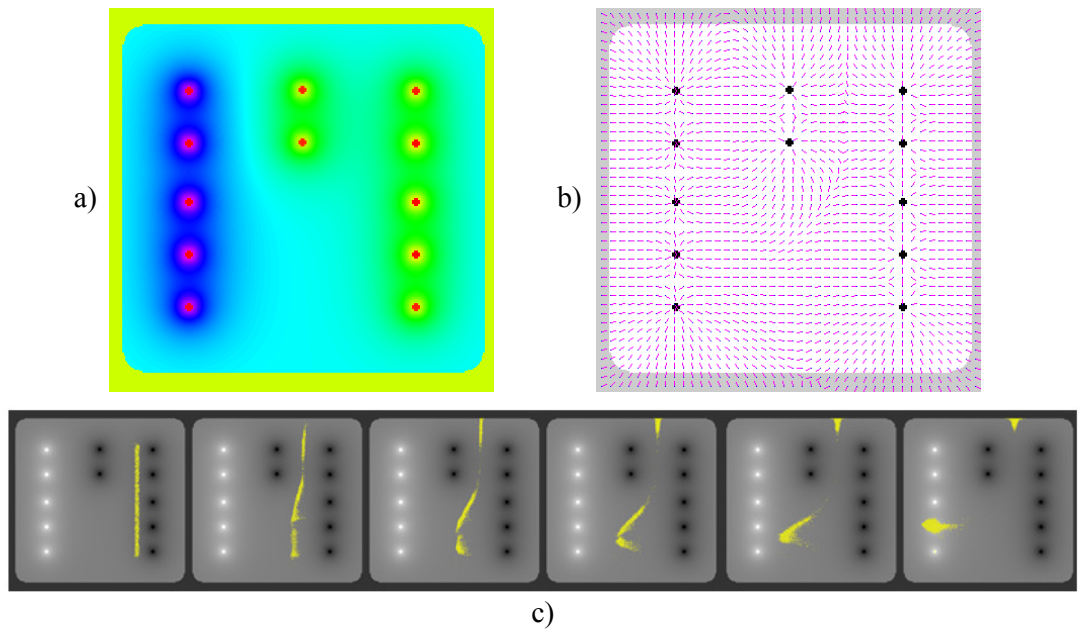

**Figure 12:** Particle approximation of obstacle avoidance during electrotaxis.
(a) 2D representation of current gradient between current sources (middle and right) and sinks (left)
(b) Vector field representation of current gradient field
(c) Migration of particle collective towards current sink electrodes avoiding obstacles. Image snapshots taken at 225, 275, 310, 350, 390 and 495 scheduler steps (bottom)

CONCLUSION AND DISCUSSION

We instigated prototype experiments in bottom-up routing and circuit pattern formation based upon observations of the self-assembly of efficient protoplasmic transport networks by the true slime mould *Physarum polycephalum*. *Physarum* is a *living* computing substrate capable of complex morphological adaptation. We interfaced the *Physarum* plasmodium with Mills' Extended Analog Computer, a spatially implemented analog computing substrate in an effort to effect programmable control of the plasmodium migration. We found that the direction, migration and obstacle avoidance behaviour of *Physarum* slime mould could be guided by the spatial distribution of electric currents formed on the EAC substrate. Previously the behavioural control of slime mould has been achieved by using white light (Hader & Schreckenbach, 1984), chemical gradients (Knowles & Carlile, 1978), and temperature gradients (Wolf, Niemuth, & Sauer, 1997). Although the negative phototactic behaviour was found 60 years ago, this is the first to show that the *Physarum* cell can be controlled electrically.

Conventionally, circuit design is embodied as a top-down process where a pre-stored template of the desired circuit is used to enforce the structure on the circuit substrate (for example by photo-lithography combined with chemical etching and deposition processes). Although this results in highly miniaturised circuits, such designs are fixed. Alternative approaches to allow reconfigurable circuits, such as electrically programmable read only memory and field-programmable gate arrays, exist but the new configurations are typically completely overwritten as pre-defined patterns in place of the previous configuration (apart from notable exceptions such as (Thompson, 1997)).

To enable concrete implementation as circuits, transport networks formed by *Physarum* must be converted into electrical circuit patterns. By performing image analysis to extract the tube network and remove artefacts from the growth substrate the network pattern may then be deposited by conventional means. Alternatively it may be possible to exploit the inherent nutrient transport function of the tube network to render the network pattern by feeding the plasmodium with particles of conducting material, such as microscopic gold beads. It has previously been demonstrated that the protoplasmic streaming within the tube network may be harnessed to implement controlled transport and mixing of external substances (A. Adamatzky, 2010a). The network pattern may be fixed, and the final circuit recovered, by inducing the conversion from plasmodium stage to the hibernating sclerotium stage.

However the generation of a fixed circuit pattern does not fully exploit the repertoire of behaviour of the slime mould. It is possible that plasmodium may be regenerated from the sclerotium stage by a careful re-introduction of moisture and nutrients, thus rendering it suitable for re-use or continuous circuit evolution. The low-level local stimuli which guide reconfiguration of the *Physarum* plasmodium enables reconfigurable circuit patterns to allow modification to the network structure in response to changing conditions or requirements. Guided low-level plasmodium routing may also mimic neural circuit evolution, where modification of the connectivity structure is a gradual, iterative, and selective process. Indeed the auto-catalytic properties of the reinforcement of protoplasmic tube network of the *Physarum* plasmodium has been shown to be equivalent to a memristive reinforcement learning process (Pershin, La Fontaine, & Di Ventra, 2009).

In this early investigation into dynamical routing the coupling of the plasmodium to the configuration mechanism of the EAC is a one-way process. We have not considered any effect that the migrating plasmodium has on the EAC conductive sheet. Since the electrode array can be used to both input and read the current distribution it is feasible that the current spatial pattern of the plasmodium could be fed back to the configuration mechanism. This information could be used to further control the transport network (circuit) pattern, to selectively reinforce or diminish particular parts of the circuit by adjusting local sink or source currents. A method based on evolutionary algorithms to optimise a pattern recognition task was implemented by (Ainslie et al., 2002) using the EAC. A neural network optical feedback method to control plasmodium shape within a confined well was implemented by (Aono, Hara, Aihara, & Munakata) for combinatorial optimisation problems. A feedback method to dynamically influence and guide the evolution of 2D spatial virtual transport networks by the particle model employed in this paper was given in (J. Jones, 2012). These methods suggest that it may be possible to couple the complex spatial evolution of biological transport networks to the exploit efficiencies in classical conventional digital computers and spatially represented analog computers alike.

ACKNOWLEDGEMENT

The work was partially supported by the Leverhulme Trust research grant F/00577/1 ``Mould intelligence: biological amorphous robots''.